\documentclass[sigconf]{acmart}
\renewcommand\footnotetextcopyrightpermission[1]{} 
\usepackage{booktabs} 
\usepackage{listings}
\usepackage{color}
\usepackage{xspace}
\usepackage{tikz}
\usetikzlibrary{positioning}
\tikzstyle{block} = [draw, thick, rectangle, rounded corners]
\usepackage{caption}
\definecolor{dkgreen}{rgb}{0,0.6,0}
\definecolor{gray}{rgb}{0.5,0.5,0.5}
\definecolor{mauve}{rgb}{0.58,0,0.82}

\newcommand{\etal}{\emph{et al.}\xspace}

\setcopyright{none}

\acmDOI{10.475/123_4}

\acmISBN{123-4567-24-567/08/06}

\usepackage{xspace}
\usepackage{color}
\usepackage{subcaption}

\usepackage{enumitem}
\newlist{compactitem}{itemize}{3} 
\setlist[compactitem]{label=\textbullet, leftmargin=1em, labelindent=0.1em, itemsep=0em, parsep=0em}

\usepackage{sutton_math_symbols}

\newcommand{\theAcro}{\textsc{Craic}\xspace}
\newcommand{\theAcronym}{Cleaner of Repetitive Areas In Comments}

\begin{document}
\title{Deep Learning to Detect Redundant Method Comments}


\author{Annie Louis}
\affiliation{%
  \institution{University of Edinburgh}
  \city{Edinburgh, UK} 
  \postcode{EH8 9AB}
}
\email{alouis@inf.ed.ac.uk}

\author{Santanu Kumar Dash}
\affiliation{%
  \institution{University College London}
  \city{London, UK} 
  \postcode{WC1E 6BT}
}
\email{santanu.dash@ucl.ac.uk}

\author{Earl T. Barr}
\affiliation{%
  \institution{University College London}
  \city{London, UK} 
  \postcode{WC1E 6BT}
}
\email{e.barr@ucl.ac.uk}

\author{Charles Sutton}
\affiliation{%
  \institution{University of Edinburgh}
  \city{Edinburgh, UK} 
  \postcode{EH8 9AB}
}
\email{csutton@inf.ed.ac.uk}


\begin{abstract}

Comments in software are critical for  maintenance and reuse. But
apart from prescriptive advice, there is little practical support or quantitative understanding
of what makes a comment useful. In this paper, we introduce the task of
identifying comments which are uninformative about
 the code they are meant to document. To address this problem, 
we introduce the  
notion of comment entailment from code, high entailment  
indicating that a comment's natural 
language semantics can be inferred directly from the 
code. 
Although not all entailed comments are low quality,
comments that are too easily inferred, for example,
comments that restate the code, are widely discouraged
by authorities on software style.
Based on this, we develop
a tool called \theAcro which scores method-level comments for redundancy. Highly redundant 
comments can then be expanded or alternately
removed by the developer. 
\theAcro uses deep language models to exploit large software corpora 
without requiring expensive manual annotations of entailment. 
We show that \theAcro can perform the comment entailment task with 
good agreement with human judgements. 
Our findings also have 
implications for documentation tools. For example, we
find that common tags in Javadoc are at least \emph{two
times more predictable}  from code than
non-Javadoc sentences, suggesting that Javadoc tags 
are less informative than more free-form comments.


\end{abstract}

%
%



\maketitle

\section{Introduction}

Reading code is central to software  maintenance.  
Studies have suggested that
programmers spend as much or more time reading and browsing code 
as actually writing it \cite{swebok:2004,latoza06maintaining,ko06exploratory}.
Naturally, developers are advised to 
write code so that it is easier to read
later, perhaps most famously by Knuth: ``Instead of imagining that our main task is 
to instruct a computer what to do, let us concentrate rather on explaining to 
human beings what we want a computer to do.'' ~\cite{knuth1984literate}.
But it is not easy to write code that is easily read: Code cannot always 
be made self-explanatory using descriptive names, and in any case, 
code cannot explain why the current approach was taken and others were not
\cite{raskin}. In such cases, comments are an important addition to enable understanding
code.
While comments play many roles~\cite{Pascarella:2017,Jml,
Cheon02aruntime}, an important role is to explain and clarify code.
Such comments have been called purpose, or \emph{explanatory comments} 
\cite{Pascarella:2017},
and are the focus of our work.


But not all explanatory comments are equally useful, and
writing useful comments requires experience and judgement. 
For example, developers are advised by no less an authority
than SWEBOK that ``some comments
are good, some are not'' \cite{swebok:2004}.
Advice from Google, as part of a carefully curated series of articles
that is distributed to all Google engineers,
explicitly encourages developers that
``comments are not always good'' and specifically to
``avoid comments that just repeat what the code does''.~\footnote{\url{https://testing.googleblog.com/2017/07/code-health-to-comment-or-not-to-comment.html}} Jef Raskin \cite{raskin} discourages
inline comments that are ``problematical... useless'' 
because they are redundant.
Finally, another authoritative guide on programming style states:
``Good comments don't repeat the code or explain it. They clarify its intent. Comments should explain, at a higher level of abstraction than the code, what you're trying to do''  \cite{mcconnell2004code}.
This advice motivates a research problem, which is only beginning to receive attention in the software engineering literature: 
How can we support developers in writing more useful
 explanatory comments? As a first step, in this paper
we ask: How can we \emph{discourage}
developers from writing explanatory comments that are ``useless'' 
because they ``just repeat what the code does''?

To an academic software engineering researcher, the expert advice that we have cited may seem counterintuitive,
perhaps even self-contradictory: If explanatory comments exist in order to explain the code,
isn't it necessary that, at least to some extent, they repeat what the code does?
It seems difficult to reconcile this natural line of reasoning against that seemingly contradictory
advice above from authorities on software development. 

To resolve this dilemma, consider two examples of real-world
Java methods.
In \autoref{fig:ex-redundant}, the comment contains nothing more
than two identical restatements of the method signature.
The code would contain
exactly the same information if the comment were deleted entirely.
The same is true of the Javadoc documentation --- there is no situation
in Java documentation or development in which a method comment is visible
but not the method signature.
Contrast this comment with the one in 
\autoref{fig:ex-informative}. While the first and final sentences are simple
restatements of the signature, the other two sentences explain the effect
of the parameter setting in more detail.  We argue that \autoref{fig:ex-redundant}
is the type of explanatory comment that is ``just restating'' that the authorities
intend to discourage. 
But in our corpus of popular open-source Java projects on Github,
we find that such
redundant comments are prevalent  (\autoref{sec:experiments}). 
Based on our own experience, we suggest a few hypotheses as to why such comments
are discouraged. First, they clutter a codebase, making it difficult for
a reader to find important logic. Second, they can pose a maintenance burden
as the code changes \cite{spinellis10}.
Finally, and insidiously, they can trick a programmer into
thinking that the code is well explained, when in fact design rationales and deeper
facts about the code, such as the ones in \autoref{fig:ex-informative},
are missing.

\begin{lstlisting}[float=t, xleftmargin=4mm, label=fig:ex-redundant,caption={A method and comment from the \texttt{liferay-plugins} project. This comment simply restates the method name.},belowskip=-0.6 \baselineskip]
/* Returns the projects entry persistence. 
   @return the projects entry persistence  */
public ProjectsEntryPersistence getProjectsEntryPersistence() { 
  return projectsEntryPersistence; 
}
\end{lstlisting}

\begin{lstlisting}[float=t, xleftmargin=4mm, label=fig:ex-informative, caption={A method and comment from the \texttt{aws-sdk-for-android} project. This comment explains
  how to use the method.},belowskip=-0.7 \baselineskip]
/* Returns the minimum part size for upload parts. Decreasing the minimum part size
 causes multipart uploads to be split into a larger number of smaller parts. Setting 
this value too low has a negative effect on transfer speeds, causing extra latency 
and network communication for each part. 
@return The minimum part size for upload parts. */
public long getMinimumUploadPartSize() { 
  return minimumUploadPartSize; 
}
\end{lstlisting}

In this paper, we introduce a machine learning (ML) framework to help developers write better,
more informative comments.
Our framework aims to identify which explanatory comments provide the most information about
the code, by exploiting the fundamental insight from information theory that sentences
that are highly predictable provide little new information. In the extreme case, 
comments that are too easily inferred from the code might trivially restate 
what the code does, or even trivially restate the method signature, as
in \autoref{fig:ex-redundant}.
By highlighting explanatory
comments that are easily inferred from the code, we hope to encourage developers to write
better comments. A developer who sees all her comment sentences
marked as easily inferred, e.g. during a code review, might be motivated
to write in more detail --- for example,
to write good explanatory comments
that cover design decisions or that ``explain why the program is being written, and the rationale for choosing this or that method'' 
\cite{raskin}.

More generally, we introduce a new research problem which
we call \emph{comment entailment}, named
after the well-studied problem of textual entailment \cite{2013Dagan} 
in the Natural Language Processing (NLP) literature.
The comment entailment problem is to determine which sentences
in a comment logically follow from the information in the code.
This is a general problem which we hope will have many uses within software engineering;
we suggest that the categorization of comments as entailed versus non-entailed, essentially
``does the comment describe the content of the code,'' is a fundamental
axis along which to characterize comments. Many entailed comments are high quality;
for example, summary comments, which briefly explain the purpose of a 
 method or class,  are often seen as desirable \cite{mcconnell2004code}. 
 But comments that are too easily inferred, such as those that are  ``word-for-word'' restatements of the code,
as in \autoref{fig:ex-redundant}, do not provide any additional
explanatory power and are discouraged by the authorities cited above.

The entailment problem is
technically challenging because it 
requires a \emph{bimodal software analysis} that considers both the source code
and the natural language comment simultaneously.
It is perhaps for this reason that there 
is almost no support in popular development environments to help developers write better explanatory comments.
We exploit recent advances in deep learning in NLP
 to develop a model that scores the level of 
redundancy in the comment. 
We present a first approach to the comment entailment problem
based on applying deep \emph{sequence-to-sequence (seq2seq) learning}  \cite{sutskever14,bahdanau2014neural}.
Comment sentences that have
\emph{highest} average probability conditioned on the code are those that
we deem as potentially low quality, under the rationale that
they are easily inferred.

We incorporate these models into \theAcro\footnote{\theAcronym, pronounced
``crack''.}, a tool to detect uninformative comments and guide software
developers to write better comments.  
Currently, \theAcro provides a textual ranked list of the most predictable
comments or optionally removes them entirely if
desired. 
Future versions could integrate into workflows for code review,
for example, colouring comments in a
heatmap, in a manner analogous to Tarantula \cite{tarantula}, to show a reviewer
which regions of
the code contain highest concentrations of uninformative comments.
\theAcro promises to gradually
improve the quality of a codebase's comments by nudging developers away from
uninformative comments and toward deeper, more useful explanations.


Our main contributions are:
(a) We introduce the comment entailment problem
(\autoref{sec:problem});
(b) We introduce a first approach for this problem based on sequence-to-sequence learning 
(\autoref{sec:deep});
(c) We show that sequence-to-sequence methods are effective
at predicting comments, as measured by perplexity,
i.e. they are capable of using code to improve
comment prediction, compared to a unimodal 
language model trained only on comments  (\autoref{sec:results});
(d) We present evidence
that \theAcro effectively identifies
redundant comment sentences, 
correlating strongly with human judgements of entailment (\autoref{sec:annotation}), based
on a new data set of code-comment pairs that we have collected;
(e) Finally, we explore how our framework can be used to make recommendations about the
design of documentation tools. We examine the hypothesis
that in Java, some kinds of Javadoc comments, such as in \autoref{fig:ex-redundant},
 are often uninformative. We indeed find that the predictability of sentences in common Javadoc 
fields is more than two times higher than  non-Javadoc sentences
  (\autoref{sec:javadoc}).
Our data, trained models and software are available at \emph{[link removed for blind submission].}




\section{Related work}

Software is bimodal:  it combines an algorithmic channel, that targets devices,
and a natural language channel, comprising comments and identifiers, aimed at
developers.  With the early and notable exception of literate
programming~\cite{knuth1992literate}, most research has focused on one of the
two channels in isolation.  As \autoref{sec:problem} makes clear, \theAcro
targets the bimodal problem of deciding whether a method entails a comment.

Our survey of relevant work in the software engineering community begins by
discussing related work focused solely on software's natural language channel,
then moves to more closely related bimodal work.  
Bimodal analysis is also motivated by
a growing line of work that applies data mining and ML techniques to software
repositories, especially work on language models
for code, which we review in more detail.

\textbf{NL channel (unimodal)}: Researchers have developed unimodal methods to
study both comments and names in software. For example,
Binkley and colleagues have measured the comprehensibility of identifier
names~\cite{binkley2011improving}.  Researchers have customised part of speech
tagging for tokenised identifier names and have mined semantically related word
pairs, by mapping the main action verb of a function's header comment to the
main action verb in its signature~\cite{gupta2013part}.  
Also relevant to our work is
a classification created by~\citet{Pascarella:2017} for Java comments. 
This work annotates a corpus of Java comments, placing comments into
categories such as those explaining the purpose, those 
informing how to use the code, providing metadata, license information, 
or \emph{todo} signals. The aim is to create a comprehensive classification
of the role of comments in software. This work also develops a Naive Bayes classifier
which predicts the category for a comment. In \theAcro, we seek
to identify redundant comments for removal to improve the 
readability of a codebase. In \autoref{sec:categories}, we examine how our predictions
relate to the categories identified by \citet{Pascarella:2017}.

\textbf{NL+code channels (bimodal)}: 
There has been some work on developing bimodal methods for the NL and code
channels of software. \citet{khamis10automatic} studied both the quality of
the NL comments, and the consistency between code and the comments to assess
quality of comments. \citet{steidl13quality} classified comments using machine
learning using four features: consistency of the comment, coherence of code and
comments, completeness of comments at focal points in the code, and usefulness of
the comment in describing the code. \citeauthor{ibrahim2012relationship} studied how to identify
code changes that trigger comment
changes~\cite{ibrahim2012relationship}.  Fluri \etal used lexical similarity and
heuristics to connect comments to code~\cite{fluri2007code}.  CloCom extracts
commented code from a codebase, then finds uncommented clones using detection as
a black box~\cite{wong2015clocom}. 

Tan and coauthors have presented methods for automatically producing code
annotations from comments~\cite{icomment,tan2011acomment, tcomment}. First, iComment~
\cite{icomment} looks at comments to extract rules that should govern the code
and then verifies whether the code accompanying the comment obeys the rules.
aComment extracts assertion macros from code, and assertional phrases from comments, and combines
them~\cite{tan2011acomment}. \citet{tcomment} analyses Javadoc comments to infer
properties of the method accompanying the comment. It then generates random test cases for the code to
identify inconsistencies between the comment and the code.
Our work on \theAcro is complementary to
aComment and iComment. Indeed, the sentences that are entailed from the code
are, in many cases, likely to be explanatory sentences rather than sentences
that make assertions as considered in the previous work, in other words,
precisely those sentences that were labelled as ``unexploitable'' by
\citet{padioleau09listening}; part of our goal in \theAcro is to guide
developers to make those explanatory, unexploitable sentences better.

Software traceability (see \cite{cleland2012software} for some recent work) is
also an inherently multimodal problem, but in a different way than the bimodal
problems we consider here, as names, comments, and code are embedded within the
same files, rather than requiring inference of cross-document links as in
traceability. \citet{miltos-bimodal} also consider a different type of bimodal
problem in software, presenting a model for code search from natural language
queries.

Closely related to our work is \citet{movshovitz13natural} who
develop topic models that jointly model comments and code; however, this work
focused only on autosuggestion rather than comment entailment.  Also,
sequence-to-sequence models have been applied within software engineering to the
API mining problem \cite{Gu2016deepapi}. But this work also did not consider the
comment entailment problem. Instead it predicts code based on natural language,
rather than comments based on code.  More recently, \citet{LinWPVZE2017:TR}
apply sequence-to-sequence learning to program synthesis from natural language,
again an inverse of the comment entailment problem that we propose.  Finally,
interesting recent work \cite{oda15,fudaba15} generates psuedo-code, which can
be viewed as a type of comment, from code using machine translation.
Unfortunately, the pseudo-code that can currently be generated by these methods
seems to be relatively literal transcriptions of the code, line-by-line.

\textbf{Language Models for code (unimodal)}: \citet{hindle12naturalness} were the first to
apply $n$-gram language models (LM) to source code.  \citet{allamanis13mining}
continued in this line by presenting the first source code LM trained on over a
billion tokens and demonstrating that predicting identifier names causes the
most difficulty to current LMs for code.  LMs for code have been applied widely,
to discover syntax errors \cite{campbell14}, to learn coding conventions
\cite{naturalize}, and within cross-language porting tools
\cite{nguyen13lexical}.  Many language models that are specifically adapted to
code have also been proposed
\cite{nguyen13statistical,maddison2014,hellendoorn17deep,raychev16,bielik16,DBLP:journals/corr/AmodioCR17}.
Recent work has also applied deep language models for code in a unimodal
fashion.  Feedforward neural network language models, simpler than the recurrent
models applied here, have been applied to code by \citet{neural-naturalize}.
Deep language models, such as RNNs and LSTMs, have been presented for the
unimodal setting of code by several authors \cite{white15deep,dam16lstm}.


\section{Problem Definition}
\label{sec:problem}

As a way of attacking the problem of identifying redundant comments, in this
paper we define and address a task which we call \emph{comment entailment}. 
Intuitively, the comment entailment problem is to identify whether a snippet of
code logically implies the statements made by a natural language comment.
Identifying entailment is easier than directly identifying redundant
uninformative comments because entailment specifies how to exploit code to
decide whether a comment is uninformative.  Our definition will not be entirely
formal, because, although the semantics of code can be described formally, the
full semantics of natural language is beyond the reach of current attempts at
logical formalization.

The comment entailment problem considers as input a snippet of source code $M$,
such as a block, method, or class, and a natural language sentence $C$ from a
comment that is associated in the code with $M$.  For brevity we will call $(M,
C)$ a \emph{code-comment pair}.  $M$ might be a single method in Java, and $C$
a sentence from the method-initial Javadoc comment.  For the purpose of the
comment entailment problem, we assume that we know in advance that $C$ is
intended to comment on $M$; in many cases, such as class- and method-level
comments, detecting which comments are intended to describe which regions of
code can be performed accurately using simple heuristics.  The entailment
problem is defined at the level of a sentence in a comment, rather than the
comment as a whole, because comments vary considerably in length. There are
many examples, such as \autoref{lst:getRegId}, where a longer comment will
contain both some sentences that are entailed and some that are not. Therefore,
a sentence-level notion seems more useful.

The {comment sentence} $C$ \emph{is entailed by} the code snippet $M$, which we
denote $M \Rightarrow C$, if the content of the
text $C$ can be semantically inferred by a reader solely from information internal to $M$.
Here ``semantically inferred'' means that a developer
can verify that the sentence $C$ is true, based solely on the method $M$.
This definition depends on what information is known by the reader, for example,
an expert programmer may understand many details about the project, the language,
 the standard library, and so on,
that allow her to verify comment sentences that a novice programmer cannot. 
This dependence cannot be fully removed, because whether a piece
of writing is clear always depends on its audience.
However, we claim that there is a core knowledge
shared by professional programmers of a language that allows them to consistent
judgement whether a comment correctly describes a method. We provide evidence for this claim by measuring
interannotator agreement in \ref{sec:annotation}.


\begin{lstlisting}[float=t, caption={This listing from the \texttt{android} project 
contains three code-comment pairs: the first sentence is completely entailed, 
the second is not, since it depends on the semantics of getString, while the third is partially entailed.} , label=lst:getRegId,belowskip=-0.8 \baselineskip]
/**
 * Return the current registration id.
 * If result is empty, the registration has failed.
 * @return registration id, or empty string if the registration is not complete.
 */
public static String getRegistrationId(Context context) { 
  final SharedPreferences prefs = context.getSharedPreferences(PREFERENCE, Context.MODE_PRIVATE); 
  String registrationId = prefs.getString(``dm_registration'',''); 
  return registrationId; 
}
\end{lstlisting}

\autoref{lst:getRegId} shows an example taken from the \texttt{android} project
that has a comment containing three sentences: on line 2 (Sentence $C_1$), line
3 ($C_2$), and line 4 ($C_3$). By definition, this example contains three
code-comment pairs.  Each pair represents a different entailment relation.
Sentence $C_1$ is completely entailed by the method, as it is simply a
restatement of the method name.  $C_2$, on the other hand, depends on the
semantics of the {\tt prefs.getString} method and hence not is directly
entailed by $M$. Finally, $C_3$ is partially entailed: the empty result
assertion is not immediate.

Instead of logical yes/no definition of
entailment, we suggest a more flexible notion of an entailment score
 $S(M \Rightarrow C)$, which is a real number that measures the degree
 of entailment. We will take the convention that lower scores indicate
a higher degree of entailment. This allows us to produce
a ranked list of comment sentences as
more or less entailed.

It is important to clarify the implications of comment entailment.
We \emph{are not} claiming that entailed comments are bad, nor are we
claiming that non-entailed comments are good. 
To the contrary, both of these incorrect statements have clear counterexamples.
A summary comment that briefly explains the algorithm in a large method
is an entailed comment that is often considered good \cite{mcconnell2004code}.
Conversely, a completely unrelated comment, such as a comment from the Linux
kernel pasted above the function in \autoref{lst:getRegId} is a non-entailed
comment which is clearly bad. 
Instead, we make two claims. The first is that entailed versus non-entailment
is a conceptually useful distinction. For example, good entailed comments (among other roles)
describe the code
at a higher level of abstraction, as recommended by \citet{mcconnell2004code},
and good non-entailed comments (among other roles) can explain rationale,
as recommended by \citet{raskin}.
Secondly, entailed comments that are too easily inferred
are not useful, and should be discouraged. At the very least, 
if \emph{all comments} in a file are easily inferred, then the comments
are likely to be missing important information such as design rationales.
%

\section{Deep Learning Comment Entailment}
\label{sec:deep}

%
%
One approach to comment entailment  would be to use supervised learning, such as text classification, in which we train a machine learning model 
to predict a binary variable indicating the presence of entailment
directly from a code-comment pair.
But such an approach requires large amounts of labelled training data, in which 
programmers have annotated code-comment pairs as to whether an entailment exists,
which is time-consuming and expensive to produce.
Instead, we avoid this problem by applying machine learning
in an indirect way, which does
not require explicitly labeled examples of \emph{whether
an entailment relationship exists} for training.\footnote{We will still require a small
amount of labelled data to \emph{evaluate} the model,
but this is much less of a concern, as long
experience in the machine learning community has shown
that the amount of data required to evaluate
the model can be several orders of magnitude
smaller than the amount of data
required for training.}

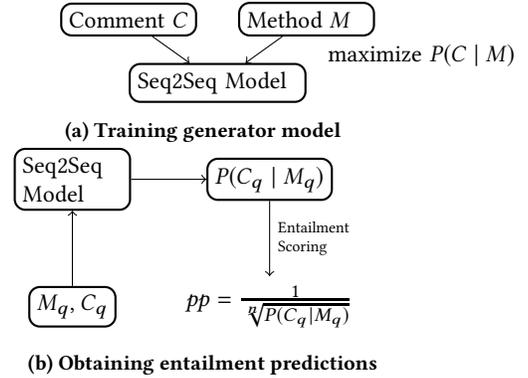
\begin{figure}
	\centering
	\begin{subfigure}{.45\linewidth}\centering
		\begin{tikzpicture}[node distance=0.5cm]
		\node[block, text width = {width("Seq2Seq Model") + 5pt}](obj){Seq2Seq Model};
		\node[block, above = 10pt and -10pt of obj.north east, text width = 5em ] (i_m){Method $M$};
		\node[block, above = 10pt and 10pt of obj.north west, text width = 5em] (i_c){Comment $C$};
		\draw[->] (i_m) -> (obj);
		\draw[->] (i_c) -> (obj);
		\node[above right = 3pt and 5pt of obj.east, text width = {width("maximize $P(C\mid M)$") + 5pt}](explain){maximize $P(C\mid M)$};
		\end{tikzpicture}
		\caption{Training generator model}
		\label{subfig:}
	\end{subfigure} 
	\begin{subfigure}{\linewidth}\centering
		\begin{tikzpicture}
		\node[block, text width = {width("generator") + 2pt}](obj){Seq2Seq Model};
		\node[block, below = of obj] (i_m){$M_q$, $C_q$};
		\node[block,right = of obj] (i_c){$P(C_q\mid M_q)$};
		\node[below = of i_c] (pp){$pp = \frac{1}{\sqrt[n]{P(C_q\mid M_q)}}$};
		\draw[->] (i_m) -> (obj);
		\draw[->] (obj) -> (i_c);
		\draw[->] (i_c) -- (pp) node [midway, right, text width = width("Entailment"), font=\scriptsize] (TextNode) {Entailment Scoring};
		\end{tikzpicture}
		\caption{Obtaining entailment predictions}
		\label{subfig:ppx}
	\end{subfigure}
	\caption{Training the comment generator model and using it to compute entailment scores. Here $pp$ stands for perplexity and $n$ is the number of tokens in $C_q$.}
	\label{fig:craic_overview}
\end{figure}

Our approach is based on language modelling.
An overview of our approach can be seen in \autoref{fig:craic_overview}.
 It consists of two stages.
First, we train recurrent neural network language models to \emph{generate}
comments based on code (\autoref{sec:seq2seq}), by which we mean that they define a probability distribution
$P(C|M)$ over comment sentences given code. We use deep learning methods 
because they are currently the most effective language models
for natural language text like comments.
Second, once we have such a model, we can use the probability values to measure the
 predictability of the comment sentence.
Comments that are too easy to predict by the model (conditioned on the code) 
are likely to be easily inferred by developers as well, 
and hence less informative. We use the probability $P(C|M)$ to define
a numerical score called perplexity (\autoref{sec:entailment}). 
Comment
sentences with low perplexity are most easily predictable.
Note
that for this approach, we only need a collection of code snippets paired
with comments written for them, which is readily available
from open source code bases, without requiring us to collect large amounts of
explicit annotations of entailment
decisions. 
 We will show that despite this lack of explicit supervision, these scores
correlate with human judgements (\autoref{sec:annotation}).


\subsection{Deep Sequence-to-Sequence Learning}
\label{sec:seq2seq}

Now we describe the sequence-to-
sequence learning framework that underlies our method. 
First, a language model is a probability distribution over strings.
Using the chain rule of probability, we can write
the probability of a sequence $w_1\ldots w_N$ of words as
\begin{equation*}
P(w_1 \ldots w_N) = P(w_1)P(w_2|w_1) \cdots P(w_N|w_1 \ldots w_{N-1}).
\end{equation*}
Different language models approximate in different ways the individual terms
$P(w_i|w_1 \ldots w_{i-1})$ in this product.
The earliest types of language models that were
applied to source code were $n$-gram models,
which make the Markov assumption that $n-1$ previous tokens
of context are sufficient for predicting each word.
Standard $n$-gram models do not perform as well as the deep language
models that we describe next, although for code, the extension of
cache language models \cite{localness-software,hellendoorn2017fse} are considerably better, 
and competitive with deep models. Nevertheless, even these cache models
are not easily applied to the sequence-to-sequence learning setting that we require here,
so we do not consider them further.

Deep language models remove the Markov assumption, achieving better performance
than traditional $n$-gram 
language models. The current state of the art
in natural language \cite{mellis17lstm} are language models
based on a type of recurrent neural network (RNN)
called the long-short term memory network (LSTM) \cite{lstm}.
More details on the LSTM can be found in \citet{Goodfellow:book}, but at a high level,
an LSTM computes a hidden state vector $\hB_i \in \R^K$,
that corresponds to every word $w_i$ and that
summarizes the information from words $w_1 \ldots w_{i-1}$ in the sequence. 
The size of the hidden layer $K$ is a parameter that we set during development.
An LSTM language model
computes the probability of a sequence
based on two neural networks $f_\mathit{lstm} : \R^K \rightarrow \R^K$
and $f_\mathit{vocab} : \R^K \rightarrow [0,1]^V$, where $V$ is
the number of words in the vocabulary. The probability of a sequence
is computed as
\begin{align}
	\hB_i &= f_\mathit{lstm} (\hB_{i-1}, w_{i-1}) 	 \label{eq:f_lstm}\\
	P(w_i | w_1 \ldots w_{i-1}) &= f_\mathit{vocab} (\hB_i). \label{eq:f_vocab}
\end{align}
Here the function $f_\mathit{lstm}$ is an ``LSTM cell'', which
is essentially a neural network that computes the next
hidden state from the previous ones, and includes
several so-called ``gates'', such as the forget gate
and the output gate. The function $f_\mathit{vocab}$ is 
a feedforward neural network that computes
a distribution over output words $w_i$ given the current value
of the hidden state $\hB_i$.

The language models we just described are unimodal and trained on strings from 
one language. But 
often, we wish to predict one sequence
from another one; for example, given a 
sentence written in French, we might 
wish to translate it into English. Sequence to sequence models
are built to learn such mappings between sequences in two languages \cite{sutskever14,bahdanau2014neural}, 
and are currently the state-of-the-art for machine translation. They work as follows.
Suppose we want to predict an output sequence 
$w_1 \ldots w_N$ given an input sequence
$v_1 \ldots v_M.$ First we run an LSTM
on $v_1 \ldots v_M$ to compute a final hidden
state $\hB_M$ by iterating \eqref{eq:f_lstm}. Then, to define a distribution
 $P(w_1 \ldots w_N | v_1 \ldots v_M)$,
 we use a second LSTM again following equations \eqref{eq:f_lstm}
 and \eqref{eq:f_vocab}, where the initial state
 of the second LSTM is  $\hB_M$. By reusing the initial state
 in this way, we train the first LSTM so as to summarize
 the information from the first sequence that is relevant
 to the second.
The parameters of both of the LSTMs are jointly
 trained by gradient descent to maximize
 $P(w_1 \ldots w_N | v_1 \ldots v_M)$.

It is this sequence to sequence learning model which is useful for 
generating comments conditioned on code $P(M|C)$, as we explain in the
next section.



\subsection{\theAcro: Deep Sequence Models for Comment Entailment}
\label{sec:entailment}

Now we can describe how LSTM language models
and sequence-to-sequence learning can be applied
to develop a method for the comment entailment
problem. 
In both cases, the entailment score is a measure 
of predictability or probability of a comment sentence under the 
respective language model.

In a unimodel language model, we can obtain the predictability
of a comment irrespective of the code it is attached to. This probability $P(C)$ 
measures how easy it is for the language model to predict the comment. 
Note that this model
ignores the code snippet $M$. A comment will have high probability under
this model if it matches frequent word sequences seen in the training 
corpus. 
In contrast, a sequence-to-sequence model learns a distribution $P( C | M )$
for a code-comment pair $(M, C)$. Therefore, the comments
that are assigned high probability from the sequence-to-sequence
model are those that are easy to predict based on the text of $M$. We 
show in the later sections that utilizing the code $M$ results in 
a better prediction of the comment $C$. 

In fact, our tool \theAcro uses perplexity instead of probability 
(low perplexity corresponds to high 
probability) as the entailment scores.  
Given a test corpus of $n$ tokens, $w_1\ldots w_n$, the perplexity is 
\begin{equation}
pp(T) = P(w_1\ldots w_n)^{-1/n}.
\end{equation}
Perplexity is inversely proportional to the probability of the text under the model and the probability is normalized for the number of words 
in the text. 
Sentences whose
perplexity are sufficiently low are those easily predicted by the model, and hence are likely to be easily inferred
from the code. So the output of the current version of \theAcro tool is a ranked list of sentences
from the comments, lowest perplexity first.
The developer can then review those sentences, e.g. during a code review,
and  consider revising them. For example, prompted by  \theAcro, 
the developer could decide to add more design rationale, to rewrite the sentence to make it a more 
useful summary. In some cases, the appropriate choice may be to remove the redundant comment altogether;
authorities on coding style are consistent that more comments are not always better \cite{mcconnell2004code,swebok:2004}.


\section{Data}

Our entailment corpus is a large collection of Java methods paired with comments. 
We focus on method-level comments only so that we can 
draw on easily identifiable data for training our models, but our framework
can be extended to other types of comments. 
We start with a large collection of Java projects, the  GitHub Java Corpus \cite{githubCorpus2013}, 
containing 14,785 projects. 
This corpus also contains project popularity ratings compiled from the number of forks 
and watchers. We use these ratings for creating test data representative of a variety of projects. 
For easy availability of both  code and ratings, 
we use this snapshot of Github for all our experiments. 

From these projects, we identify those comments describing a
method as a whole and immediately preceding the method. We remove all other multi-line and inline comments within 
a method. The resulting corpus is a collection of pairs of method and \emph{full comment} texts.
At this point, the comment is a span of text which may contain more than one sentence. We 
call the span a \emph{full comment} to distinguish it from the single sentences (\emph{comment sentence}) that our models use.
We ignore methods that do not have comments.

This code-full comment  corpus contains over 3M pairs. 
We preprocess this 
corpus in a few ways. For code, we use a lexer to tokenize the method.
For the comments, we use the 
Stanford CoreNLP toolkit \cite{corenlp} to tokenize the text. In both methods and comments, 
we subtokenize any camelCase names into separate words. 

The resulting methods and comments can vary widely in length. 
Table \ref{tab:datastats} shows how the token counts of methods and comments
are distributed in our  corpus. The average length of a method is 75 tokens, three
times longer than the average length of a full comment. 
It is also apparent that the distribution over method lengths in the corpus is highly skewed,
with the mean length being not only larger than the median but larger even than the 3rd quartile.

We also segment the comment text into sentences, using the CoreNLP toolkit, together with additional
heuristics. For example, fields in Javadocs such as \emph{@param} and the 
accompanying parameter description are treated as a single sentence.
Once we have comment sentences, each comment sentence is paired with the method individually, resulting 
in a collection of method-\emph{comment sentence} pairs which are used in all our models. 

We randomly draw 3M method-\emph{comment sentence} pairs for training, 5000 for a validation set,
and 5000 for testing.



\section{Experiments}
\label{sec:experiments}

Here we describe how we represented the methods and comments to input into our models, model
implementation, and  the resulting performance. 
We also examine
how our best model's judgements correlate with human assessments of comment entailment, and with categories of comments
proposed by prior work.

\subsection{Input Representation}
\label{sec:compression}

As described in the previous section, our 
models are trained with pairs, each containing a method and a comment sentence. 
In contrast to comment sentences in the pair, 
method text can be arbitrarily long. This variation over method length makes training
sequence-to-sequence models rather difficult. 
Hence we developed three ways to compress a method to a maximum of $L$ tokens.

\noindent {\bf a) signature}: In some cases, the method signature alone is sufficient to
determine entailment. 
So this representation retains only the method signature and ignores the rest of the method body. 
When the signature is longer than $L$ tokens, it is truncated. 


 
\noindent {\bf b) begin-end}: Here the method is represented by a total of $L$ tokens, half taken from the start of the method and the other half from the
end. By sampling tokens from both ends, this representation makes more use of the method 
body compared to the signature based compression, such as the \texttt{return} statement (if any).

\begin{table}
\begin{tabular}{l r r}
        & {\bf Methods} & {\bf Comments}\\
Mean    &   75.18       &   28.81  \\
Median  &   30.00       &   19.00 \\\hline  
1st Quartile &    17.00    &    10.00 \\  
3rd Quartile &    72.00    &    34.00 \\  
\end{tabular}
\caption{Statistics for length (in tokens) of methods and \emph{full comments} in our 3M pairs corpus.}
\label{tab:datastats}
\end{table}

\noindent {\bf c) identifier-based}: This representation first preserves the method
signature, and then retains a subsequence of the method body comprising only salient
identifier names.  We limit the overall sequence to $L$ tokens.  While the
sequence is shorter than $L$, we incrementally add braces and names to it based
on precedence.  This precedence is over braces and names as follows: braces,
locals, globals, user-defined types, externally defined methods, locally
defined methods, and formals.  This order heuristically captures salience.  We
define locals > globals, because they name internal computations of the method;
we define externally defined methods > locally defined methods, since external
method names are often semantically significant.  Subject to $L$, we exhaust
the names in a higher salience category before moving to a lower category.
Within a category, we add names in their order of appearance in the code.
Because of $L$, braces may be empty or unbalanced, but this happens rarely
in practice.  Braces surface identifier nesting to our model.

For all three compression methods, we set $L = 50$ tokens. 
Comments are also truncated to 50 tokens.  We choose $L$ to compromise
between information and computational efficiency.

\subsection{Model Details}

During development, we examined the performance of our models with different hidden units and depth. Our best configurations of
a single hidden layer, and 2048 units for the language model and 512 units for the sequence model was used for the final 
training. We used a vocabulary size of 25000 for both our models, on both the method side and comment side. The initial learning 
rate was set to 0.5, and we used a decay factor of 0.96 which was applied when the validation perplexity did not improve over
an epoch. We used a batch size of 64 and a dropout probability of 0.65. For the language model, we truncated backpropagation at 
30 steps and use the final states of the previous batch to initalize the start state of the next batch. We use gradient descent to
optimize the models and clip the gradients at 5.0. We implemented the models in Tensorflow\footnote{ \url{https://www.tensorflow.org}}.
The language model training was done for total of 51 epochs, while the sequence models were run for 16 to 23 epochs each;
like the other hyperparameters, these values were also chosen based on performance on the validation set.

\subsection{Validation via Predictive Performance}
\label{sec:results}

In this section, we are evaluate whether our language models over comments are effective at predicting
text. In later sections, we evaluate whether the resulting perplexity scores are effective for
predicting entailment.
We evaluate two models. One is a LSTM language model trained only on comment text.
Next we examine whether sequence-to-sequence models are effective at leveraging the method code
to improve their capability to predict comments.
In these s2s models, 
we experiment with the three method summarization techniques from Section \ref{sec:compression}.
Clearly, if the sequence-to-sequence model does not display better predictive performance, then it is 
not using information from the code effectively. 
To compare comment corpus to a general English language corpus,
we also consider a state-of-the-art LSTM language model built for English newswire text
\cite{mellis17lstm}. It was trained and tested on partitions of 
the Penn Treebank \cite{marcus93building}, a corpus containing Wall Street Journal
news articles. 

Language models are typically evaluated using their perplexity on a test set, which is a collection of texts 
unseen by the 
models during training. This is standard methodology in natural language processing for measuring
the quality of a language model. Lower perplexities are better and a language model with the lowest perplexity on a test set is best at predicting strings from the language.  Previous studies of language models for
code have also reported cross-entropy, $xe(T) = n^{-1} \log_2 P(w_1\ldots w_n)$. The relationship
is therefore simply $pp(T) = 2^{-xe(T)}$.
Perplexity has an intuitive interpretation. The perplexity of a uniform distribution
over $V$ words is exactly $V$, so perplexity can be viewed as an ``effective vocabulary size'' of the model,
or how many guesses the model would need on average to predict every word in the text. The units
of measure for perplexity can be intuitively understood as ``number of vocabulary entries''. Both a
general language model, and sequence learning models can be evaluated using perplexity.

\autoref{tab:modelppx} shows the
perplexities of our models on the training, validation and test sets.
We see that indeed our language models are dramatically better at predicting comment text
than state-of-the-art models are at predicting newswire text.
The perplexity of 58 for newswire text compared to those on the order of 10 and 5 for comments.
We hypothesize that comments are easier to predict compared to natural language news, 
because comments belong to a narrower domain in terms of both
vocabulary and the productive nature of sentences. 

For the sequence-to-sequence models, we compare all three code representation 
methods from \ref{sec:compression}, namely (a) the signature-based
representation (s2s-signature in \autoref{tab:modelppx}), (b) begin-end representation (s2s-begin-end), and (c) the identifier-based
representation (s2s-identifier).
The perplexity of the best sequence-to-sequence model is about half the 
number from the language model. Hence simply
capturing the most frequent comment tokens, while informative, does not perform as well as the entailment 
models which use the method to make the predictions. In terms of which method representation is most useful, we find 
that there is an improvement upon using the method body in some compressed form (either as sampled tokens in the 
begin-end case or using identifier sequences) compared to signature only. 
Overall, this evaluation indicates 
that the sequence to sequence models are effective at predicting comments conditioned on the code, thereby providing
a proxy for entailment. 

Since s2s-begin-end and the identifier compression perform similarly, we use the simpler
begin-end model as our \emph{best model} for the rest of the analysis in this paper. 

In Table \ref{tab:ppxexamples},
 we also show qualitative examples for the highly entailed (low quality) and 
low entailment comments according to our \emph{best model}. We do not show the methods due to space constraints, but the
comments themselves are often enough to understand the distinction we are trying
to convey between redundant examples and those which would be difficult to predict from the code.

\begin{table}
\begin{tabular}{l r r r r}
              & \multicolumn{3}{c}{\bf Perplexities}\\
{\bf Model}   & {\bf Train} & {\bf Valid} & {\bf Test} \\
LM            &   7.80      & 10.34       & 9.87\\
s2s-signature &    5.70     & 6.90         & 8.26  \\
s2s-begin-end &    3.44      &  4.18       & 5.31  \\
s2s-identifier&    4.50      &  5.34       & 6.00  \\\hline
LM English newswire &  &  & 58\\ 
\end{tabular}
\caption{Performance of deep models based on their ability to predict comments. Lower perplexities are  better.}
\label{tab:modelppx}
\end{table} 

\begin{table}[h]
\begin{tabular}{|r |@{~~}p{7cm}@{~}|} \hline
      & {\bf High perplexity comments} \\
149.84 &   Such error prevents checking out and creating new branch.\\
93.50 &   Keep id to make sure temp file will be removed after use uploaded file.\\
74.36 &   Here the stacktrace serves as the main information since it has the method which was invoked causing this exception.\\
73.83 &   Assumes that statistics rows collect over time , and that none of them have disappeared.\\
73.20 &   This helps to prevent (bad) application code from accidentally holding onto extraneous garbage.\\
68.11 &    The only place this flag is used right now is in multiple page dialog icon\_style and tab\_style.\\
50.83 &   The client property dictionary is not intended to support large scale extensions to jcomponent nor should be it considered an alternative to subclassing when designing a new component. \\ \hline
 & {\bf Low perplexity comments}\\
1.03 &  setter method for ``name'' tag attribute.\\
1.07 & Finds the user id mapper with the primary key or returns null if it could not be found.\\
1.10 & Calls case xxx for each class of the model until one returns a non null result; it yields that result. \\
1.56 &  compares this uuid with the specified uuid.\\
2.84 &  @throws SettingNotFoundException thrown if a setting by the given name can't be found or the setting value is not an integer.\\
3.05 &  removes a global ban for a player. \\
3.06 &    private default constructor. \\ \hline
\end{tabular}
\caption{Example high and low perplexity comments under our \emph{best model}, and their perplexity values.}
\label{tab:ppxexamples}
\end{table}

\subsection{Comparison to Comment Categories}
\label{sec:categories}

Prior work on comments,
\citet{Pascarella:2017} has classified comments discounting their
usefulness. We examined how our model predicts comment sentences which were
involved in their manual comment classification work. 
This analysis identifies categories from the manual classification which are
deemed redundant or non-redundant by our model.
In this way, we gain intuition into the predictions of our model. We 
hypothesize that some categories of comments are more likely to be entailed than others. 
For example, comments that are categorized as explaining functionality are likely to be
more easily inferred than comments that explain the deeper rationale of the code. 

\citet{Pascarella:2017}'s corpus contains 11,226 annotated comments. 
The comments were annotated into 6 major categories: \emph{Purpose} (explain 
the functionality of the code), \emph{Notice} (warnings, alerts, and information about usage), \emph{Under
development} (todo and incomplete comments), \emph{Style} and \emph{IDE} (IDE directives and formatting
text), \emph{Metadata} (license, ownership etc), and \emph{Discarded} (noisy comments). Each category is further divided into finer sub-categories.

To compare with our work, we identified method-level comments and their code span from their corpus. 
We were able to obtain the code spans for 837 method comments successfully. For these 
comment-code pairs, we grouped the comments using the taxonomy adopted in \citet{Pascarella:2017}. Then, we studied 
how the categorisation rings with the predictions of our model.

Table \ref{tab:MSRmethodcomments} shows the categorisation of the
837 comments using the \citet{Pascarella:2017} taxonomy. Most of the method
comments belong to either \emph{Purpose} or \emph{Notice} types. This is expected as
License, Todo or Metadata comments are less unlikely to be method-level comments.
Within \emph{Purpose} and \emph{Notice} categories,  most
comments are in the \emph{Purpose-summary} (comments on what the method
does), \emph{Notice-usage} (how a method must be used, or parameter definitions)
and \emph{Purpose-rationale} (why code was written in a certain way),
and \emph{Purpose-expand} (how the code was implemented, purpose of different 
parts of the code in detail).

Our analysis shows that method level comments, unlike license or todo comments, 
are likely to fall under categories
where a notion of entailment is well-defined. This lends confidence that our 
model is trained and tested on data that is likely to have
categories like those in  Table \ref{tab:MSRmethodcomments} and entailment could 
be used to differentiate
the comments which are redundant. 

We used our best model (s2s-begin-end) to examine model predictions on the above
categories of comments. 
We split each comment into sentences as our model is designed to
make predictions at the level of sentences. Each sentence is paired with the 
source code from the comment-code pair. This process gave a total
of 1352 method-comment sentence pairs for which we obtain model predictions of 
perplexity per
comment. We average the perplexities for each category and report them in Table 
\ref{tab:ppxMSRdata}.


\begin{table}
  \begin{tabular}{l r}
    {\bf category} & {\bf count}\\ 
     Purpose-summary & 716 \\
     Notice-usage    &  95 \\
     Purpose-rationale  &  14 \\
     Purpose-expand   &  8\\
     Metadata          & 3\\
     Under development-todo   &  1\\
  \end{tabular}
  \caption{The distribution of human annotated categories defined in 
\cite{Pascarella:2017} on the subset of method comments.}
  \label{tab:MSRmethodcomments}
  \end{table}

\begin{table}
  \begin{tabular}{l r  r r r}
{\bf category}	& {\bf count}	& {\bf avg} & {\bf stdev}	 & {\bf median} \\
purpose-expand   &	48	& 35.23	 & 32.64 & 23.55 \\
purpose-rationale &	25	& 20.90	 & 25.47 & 11.02 \\ 
purpose-summary   &	1152	& 19.33	 & 28.42 & 10.54 \\ 
notice-usage	    &    123	& 14.19	 & 26.93 & 5.95	 \\ 
  \end{tabular}
  \caption{Average sentence perplexities (also standard deviation and median)
 from our best model on the 1352 method-\emph{comment sentence}
    pairs from \cite{Pascarella:2017}. We ignore the infrequent metadata and under-development categories. }
  \label{tab:ppxMSRdata}
\end{table}

The \emph{Notice-usage} category represents explicit instructions on how to use a piece of code. This category
has the lowest perplexity; part of the reason is that a
lot of Javadoc is under this category, and as we show later, comments in Javadoc tags are highly predictable.
Similarly, purpose-summary is meant to say what the code does and its perplexity is 
lower compared to rationale or expand. \emph{Purpose-expand} and \emph{Purpose-rationale}
indicate how certain things in the code were done and the choices made respectively. 
These categories have highest perplexity which
matches our intuition/claims that those comments which add explanations going beyond the
code would be predicted as non-entailing.

\subsection{Comparison with Human Judgements}
\label{sec:annotation}

In this section, we validate our core claim that predictive performance from a
sequence to sequence model
can be used to develop a comment entailment method by comparing the entailment
decisions from \theAcro to the judgement of human developers.
To avoid confusion, recall that  low perplexity on 
a comment indicates that it is highly entailed by the code, high perplexity indicates a high degree of non-entailment. Now 
we verify if our scores can predict cases where human would also judge the comments similarly as entailing or not.

To perform this evaluation, we select a sample of 45 projects from the Github corpus \cite{githubCorpus2013}. This set is chosen 
such that projects of varying quality are included. For quality, we use a popularity score for each project
which is the sum of the number of forks and the number of watchers (each of the two values is first converted into z-scores before
adding them up). We then sample 15 projects from a high popularity range, 15 medium and 15 low popularity according to these scores.
In these projects, we collected methods which had maximum 100 tokens so that they would be easier for the annotators to read  without the context of the full code base.
We randomly sampled 500 method-comment sentence pairs from this set, and performed an annotation experiment. 

\subsubsection{Annotation experiment}

We hired five M.Sc. students in Informatics as annotators to provide human judgements of 
entailment. All of them have at least 2 years of experience in professional software development. 
Our interface presented a method and an associated comment sentence. The annotators read them both and
decided on one among five entailment
options:

\begin{compactitem}
\item \emph{entails}: the comment sentence is logically entailed by the method
\item \emph{does not entail}: the comment sentence is not entailed by the method
\item \emph{partly entails}: option to be used for long comment sentences where some portion of the comment sentence is entailed though
not the full sentence.
\item \emph{cannot decide}: option allows annotators to refrain from making a decision when they do not understand the
method or comment either due to high context dependence or low quality comments. 
\item \emph{un-related}: when the annotators understood the comment and the method, but were unable to see how they go together. 
\end{compactitem}

We also tracked two properties of the comments which we thought can be used to 
analyze annotator decisions based on intrinsic comment properties. Annotators could mark individual comments as \emph{incoherent} when the comment
is low quality or too short to understand. Another option allowed annotators to mark off comments which are part of \emph{javadoc}. These markings
were in addition to the entailment choice. 

We did not ask the annotators
to judge whether the comment sentence was useful, but only whether the comment was entailed by the code,
which  we argue is a more objective notion. 



Every annotator marked each of the 500 examples, but each annotator saw the examples in a different random order. 
On average, an annotator took 8 hours to complete the task.

\subsubsection{Annotator agreement}

We removed one annotator who had substantial disagreement with the rest of the annotators (measured by pairwise
Cohen's Kappa score for rater agreement). The remaining annotators had a fair level of agreement. The 
majority of the confusion was between entail and partly entail
categories. The pairwise Cohen's Kappa for the
remaining four annotators ranges from 0.1 to 0.29 indicating fair agreement. 
We noticed that there were two subgroups within
our annotators, where two of them overwhelmingly picked `entails' for ambiguous examples and two picked `partly entailed' or `not entailed'. The first pair of annotators have an agreement of 0.19 and the second 0.29.  

On one-fifth of our examples (95 out of 500), all four annotators picked the same entailment 
decision (out of \emph{five} possible) indicating that this task is meaningful
to the annotators.
A majority decision was possible on 292 examples i.e. three or all of the four annotators picked the same choice on these
examples. These numbers indicate that close to 60\% of the examples could be annotated reliably.

For our class of interest, the redundant comments, 85 samples were marked by all four annotators as ``entailed''. We will examine our model's predictions on this subset in the next section. 

Below we provide
some examples of annotator decisions. 

M1,C1 is a pair where all four annotators agreed that the comment is \emph{not entailed} by the method. 

\begin{lstlisting}[xleftmargin=4mm]
(M1,C1)
/* This method should be called before the superclass implementation.*/
public void dispatchDestroy() { 
}       
\end{lstlisting}

In M2,C2 all four annotators
agree that the comment is entailed or partly entailed by the method. 

\begin{lstlisting}[xleftmargin=4mm]
(M2,C2)
/* @throws IOexception thrown on errors while reading the matrix */
public void load(String filename) throws IOException { 
  DataInputStream dis = new DataInputStream(new FileInputStream(filename)); 
  this.in(dis); 
}        
\end{lstlisting}

M3,C3 is a pair where annotators disagreed. 

\begin{lstlisting}[xleftmargin=4mm]
(M3,C3)
/*The keyword used to specify a nullable column.*/
public String getNullColumnString() { 
  return " with null"; 
}        
\end{lstlisting}

Here, two annotators picked entailed/partly entailed and two chose the non-entailed. 
It is likely that the information about the columns being ``nullable'' is not fully inferrable
from the code (as opposed to columns containing null values already). Similarly, the return value being a \emph{keyword} is not directly 
inferrable from the code. These points may trigger a ``not entailed'' decision. At the same time,  the return type being
a string may have been considered by the two other annotators as sufficient to entail/partly entail. 

Other comments where annotators disagreed
included comments which were not fluent or were context-dependent. 
In fact, out of the 84 examples where no majority decision 
was reached, 30 were marked by annotators as \emph{incoherent}. 
Note that incoherent comment sentences also result from 
errors in the sentence segmentation performed on the comment text. 

\subsubsection{Comparison between model and human judgements}

Since our annotators could reliably annotate and agree on the examples, we now examine 
whether the entailment predictions from \theAcro match human judgements.

\begin{table}
\begin{tabular}{l r | r r r}
{\bf category}   & {\bf count} & {\bf avg} & {\bf stdev} & {\bf median}\\	
entails	         & 237	& 9.50	     & 33.23	& 2.30	\\
partly entailed	 &  12	& 14.77	     & 17.00	& 7.35	\\
not entailed	 &  39	& 115.73     &266.65	& 13.35	\\
unrelated        &   4	& 1069.73    &676.71	& 1206.36\\	
\end{tabular}
\caption{Perplexities of our best model on 292 human annotated samples with a majority agreement.}
\label{tab:annotDataPpx}
\end{table}

For this analysis, we use the 292 examples where a majority decision was reached by the
annotators. Table \ref{tab:annotDataPpx} shows the perplexities of our best model (s2s-begin-end)
on these examples split by the majority category from the annotation. 
We see that the entailed examples have lowest average and median perplexities compared to those
partly entailed, which in turn are lower than the non-entailed examples. 
This finding shows that our model predictions correspond well with
manual annotations by software developers.

Beyond agreement, it is also of interest to explore illustrative examples of when the model predictions 
are incorrect.
For the example below, all four annotators marked the comment as entailing or partly entailing but
the model assigned a high perplexity (74.3). For humans it is clear that $i$ is an index but a model
with only surface code tokens will fail to predict those comment tokens. In addition, the fact that range is constant
may be treated as subsidiary information by humans but a model score will be affected by the low predictability of these tokens. 

\begin{lstlisting}
/*construct point range (constant range) with given index.*/
public static RangePoint(int i) {
  return new PointRange(i);
}  
\end{lstlisting}

Another noteworthy example is the following 
where the model predicts the comment to have low perplexity (1.12, hence entailed). However, three out of four
annotators marked it as not entailed. 

\begin{lstlisting}[xleftmargin=4mm]
/*start and end are not primary keys, they are indexes in the result set.*/
public List<KBArticle> findByG_L (long groupId, boolean latest, int start, int end) throws SystemException {
  return findByG_L(groupId, latest, start, end, null);
}
\end{lstlisting}

Here a crucial component of the semantics of the comment depends on the negation that
is being conveyed: variables {\tt start} and {\tt end} are indexes in a result set and {\bf not} primary keys. 
A surface level sequence to sequence model is not sensitive to such nuances. 

\section{{\tt javadoc} comments}
\label{sec:javadoc}

Finally, our approach also allows us to make more systematic recommendations
about broad classes of comment sentences.  As an exemplar of this type of
study, we evaluate the usefulness of fields in {\tt javadoc} comments, to test
the hypothesis that some fields such as \texttt{@param} and \texttt{@return}
encourage developers to restate the method signature rather than providing
useful information. While these tags are useful for the document generator to
pepper the documentation with code snippets, such comments are rarely
elaborate or insightful.

To explore this hypothesis, we computed the average perplexity assigned by our
best sequence to sequence model (\emph{begin-end}) on 
the comment sentences in our 45 project dataset, a subset of which
we used for the annotation experiment. There
are 73,430 comment sentences in this set. 
We only consider those {\tt javadoc} elements that occur at least 25 times in the corpus.

\begin{table}
\begin{tabular}{l r r}
{\bf {\tt javadoc} type}  & {\bf no. sentences}  & {\bf avgppx}\\
{\bf Non-javadoc} &31688 &15.58\\ \hline
@linkplain  &  26  &15.91\\
@return     &11100 &13.28\\
@code       &1667  &10.01\\
@link       &2623  &9.56\\
@param      &18052 &6.49\\
@deprecated & 387  &6.41\\
@see        &1460  &6.13\\
@inherit    &651   &3.53\\
@throws     &5342  &2.50\\
@since      &359   &1.44\\ 
\end{tabular}
\caption{The average perplexity of sentences from different
javadoc elements and non-javadoc sentences. Low perplexity indicates more easily entailed comments. Most
 javadoc elements have much lower perplexity  than an average non-javadoc sentence.}
\label{tab:javadoc}
\end{table}

Table \ref{tab:javadoc} shows the number of sentences belonging to
{\tt javadoc} elements and the average perplexity. For comparison, the \textbf{Non-javadoc} 
of the table is the average perplexity of the sentences not belonging to a {\tt javadoc} element.
The non-javadoc sentence perplexity is around 15.
In comparison all but one of the {\tt javadoc}
element sentences have lower perplexities. Many common tags,
such as {\tt @since}, {\tt @throws}, {\tt @inherit}, 
{\tt @param} and {\tt @deprecated},
have much lower perplexity than non-Javadoc
comments, showing that these elements can be easily predicted by a simple surface analysis 
of the method body.  In fact, two of the commonly used 
fields, {\tt @param} and {\tt @throws}, are at least \emph{two times more
predictable} than
a non-javadoc comment sentence.

To validate the results from our deep models, we also
examine how {\tt javadoc} elements were treated by our annotators. Of the 500 code-comment pairs 
that were annotated, 190 were annotated by at least one of our annotators as  involving {\tt javadoc}. 
The entailment decisions on these
samples are heavily towards \emph{entailment}: 180 marked as entailing or partly entailing,  and 10 
as not entailing. 

This result has implications for the design of documentation generators like {\tt javadoc}.
It is consistent
with the claim that comments in {\tt javadoc} fields are less informative than other types of comments.
Uninformative comments
increase visual clutter and decrease readability, as evidenced by
the advice to developers cited earlier
to avoid such such comments.
Therefore, if confirmed
by more extensive studies, these
 empirical findings could motivate
an effort by designers of documentation systems to consider ways
to modify the available fields to encourage more informative comments, such as by refining
the set of available {\tt javadoc} elements, or revising the set of best practices for filling
in the existing fields.
The result also raises the possibility that that deep 
learning language models will be able to automatically generate such comment categories allowing a developer to focus
on comments that require greater knowledge and depth of understanding.

\section{Conclusion}

In this paper, we have introduced the problem
of comment entailment and used it to develop a tool to detect redundant comments.
While all entailed comments need not be of low quality, highly entailed ones, that 
can be readily detected automatically, are likely to be uninformative. 
Going forward, we aim to develop models which can identify other types of uninformative
comments with wide coverage and minimal annotation. 
Beyond comment quality, 
our entailment model could be useful in a variety of settings, for example,
as a scoring tool within code search, 
within program synthesis from natural language,
and within code summarization.

\bibliographystyle{ACM-Reference-Format}
\bibliography{comment} 


\begin{thebibliography}{00}


\ifx \showCODEN    \undefined \def \showCODEN     #1{\unskip}     \fi
\ifx \showDOI      \undefined \def \showDOI       #1{#1}\fi
\ifx \showISBNx    \undefined \def \showISBNx     #1{\unskip}     \fi
\ifx \showISBNxiii \undefined \def \showISBNxiii  #1{\unskip}     \fi
\ifx \showISSN     \undefined \def \showISSN      #1{\unskip}     \fi
\ifx \showLCCN     \undefined \def \showLCCN      #1{\unskip}     \fi
\ifx \shownote     \undefined \def \shownote      #1{#1}          \fi
\ifx \showarticletitle \undefined \def \showarticletitle #1{#1}   \fi
\ifx \showURL      \undefined \def \showURL       {\relax}        \fi
\providecommand\bibfield[2]{#2}
\providecommand\bibinfo[2]{#2}
\providecommand\natexlab[1]{#1}
\providecommand\showeprint[2][]{arXiv:#2}

\bibitem[\protect\citeauthoryear{Abran, Bourque, Dupuis, Moore, and
  Tripp}{Abran et~al\mbox{.}}{2004}]%
        {swebok:2004}
\bibfield{author}{\bibinfo{person}{Alain Abran}, \bibinfo{person}{Pierre
  Bourque}, \bibinfo{person}{Robert Dupuis}, \bibinfo{person}{James~W. Moore},
  {and} \bibinfo{person}{Leonard~L. Tripp}.} \bibinfo{year}{2004}\natexlab{}.
\newblock \bibinfo{booktitle}{{\em {Guide to the Software Engineering Body of
  Knowledge - SWEBOK}\/} (\bibinfo{edition}{2004 version} ed.)}.
\newblock \bibinfo{publisher}{IEEE Press}, \bibinfo{address}{Piscataway, NJ,
  USA}. 1--202 pages.
\newblock
\showISBNx{0769510000}
\showURL{%
\url{http://www.swebok.org/ironman/pdf/SWEBOK\_Guide\_2004.pdf}}


\bibitem[\protect\citeauthoryear{Allamanis, Barr, Bird, and Sutton}{Allamanis
  et~al\mbox{.}}{2014}]%
        {naturalize}
\bibfield{author}{\bibinfo{person}{Miltiadis Allamanis},
  \bibinfo{person}{Earl~T Barr}, \bibinfo{person}{Christian Bird}, {and}
  \bibinfo{person}{Charles Sutton}.} \bibinfo{year}{2014}\natexlab{}.
\newblock \showarticletitle{Learning Natural Coding Conventions}. In
  \bibinfo{booktitle}{{\em FSE}}.
\newblock


\bibitem[\protect\citeauthoryear{Allamanis, Barr, Bird, and Sutton}{Allamanis
  et~al\mbox{.}}{2015a}]%
        {neural-naturalize}
\bibfield{author}{\bibinfo{person}{Miltiadis Allamanis},
  \bibinfo{person}{Earl~T. Barr}, \bibinfo{person}{Christian Bird}, {and}
  \bibinfo{person}{Charles Sutton}.} \bibinfo{year}{2015}\natexlab{a}.
\newblock \showarticletitle{Suggesting Accurate Method and Class Names}. In
  \bibinfo{booktitle}{{\em Foundations of Software Engineering (FSE)}}.
\newblock


\bibitem[\protect\citeauthoryear{Allamanis and Sutton}{Allamanis and
  Sutton}{2013a}]%
        {allamanis13mining}
\bibfield{author}{\bibinfo{person}{Miltos Allamanis} {and}
  \bibinfo{person}{Charles Sutton}.} \bibinfo{year}{2013}\natexlab{a}.
\newblock \showarticletitle{Mining Source Code Repositories at Massive Scale
  using Language Modeling}. In \bibinfo{booktitle}{{\em Working Conference on
  Mining Software Repositories (MSR)}}.
\newblock


\bibitem[\protect\citeauthoryear{Allamanis and Sutton}{Allamanis and
  Sutton}{2013b}]%
        {githubCorpus2013}
\bibfield{author}{\bibinfo{person}{Miltiadis Allamanis} {and}
  \bibinfo{person}{Charles Sutton}.} \bibinfo{year}{2013}\natexlab{b}.
\newblock \showarticletitle{{Mining Source Code Repositories at Massive Scale
  using Language Modeling}}. In \bibinfo{booktitle}{{\em The 10th Working
  Conference on Mining Software Repositories}}. IEEE,
  \bibinfo{pages}{207--216}.
\newblock


\bibitem[\protect\citeauthoryear{Allamanis, Tarlow, Gordon, and Wei}{Allamanis
  et~al\mbox{.}}{2015b}]%
        {miltos-bimodal}
\bibfield{author}{\bibinfo{person}{Miltiadis Allamanis},
  \bibinfo{person}{Daniel Tarlow}, \bibinfo{person}{Andrew~D. Gordon}, {and}
  \bibinfo{person}{Yi Wei}.} \bibinfo{year}{2015}\natexlab{b}.
\newblock \showarticletitle{Bimodal Modelling of Source Code and Natural
  Language}. In \bibinfo{booktitle}{{\em International Conference in Machine
  Learning}}.
\newblock


\bibitem[\protect\citeauthoryear{Amodio, Chaudhuri, and Reps}{Amodio
  et~al\mbox{.}}{2017}]%
        {DBLP:journals/corr/AmodioCR17}
\bibfield{author}{\bibinfo{person}{Matthew Amodio}, \bibinfo{person}{Swarat
  Chaudhuri}, {and} \bibinfo{person}{Thomas~W. Reps}.}
  \bibinfo{year}{2017}\natexlab{}.
\newblock \showarticletitle{Neural Attribute Machines for Program Generation}.
\newblock \bibinfo{journal}{{\em CoRR\/}}  \bibinfo{volume}{abs/1705.09231}
  (\bibinfo{year}{2017}).
\newblock
\showURL{%
\url{http://arxiv.org/abs/1705.09231}}


\bibitem[\protect\citeauthoryear{Bahdanau, Cho, and Bengio}{Bahdanau
  et~al\mbox{.}}{2015}]%
        {bahdanau2014neural}
\bibfield{author}{\bibinfo{person}{Dzmitry Bahdanau},
  \bibinfo{person}{Kyunghyun Cho}, {and} \bibinfo{person}{Yoshua Bengio}.}
  \bibinfo{year}{2015}\natexlab{}.
\newblock \showarticletitle{Neural machine translation by jointly learning to
  align and translate}. In \bibinfo{booktitle}{{\em ICLR}}.
\newblock


\bibitem[\protect\citeauthoryear{Bielik, Raychev, and Vechev}{Bielik
  et~al\mbox{.}}{2016}]%
        {bielik16}
\bibfield{author}{\bibinfo{person}{Pavol Bielik}, \bibinfo{person}{Veselin
  Raychev}, {and} \bibinfo{person}{Martin Vechev}.}
  \bibinfo{year}{2016}\natexlab{}.
\newblock \showarticletitle{{PHOG:} Probabilistic Model for Code}. In
  \bibinfo{booktitle}{{\em ICML}}.
\newblock


\bibitem[\protect\citeauthoryear{Binkley, Hearn, and Lawrie}{Binkley
  et~al\mbox{.}}{2011}]%
        {binkley2011improving}
\bibfield{author}{\bibinfo{person}{Dave Binkley}, \bibinfo{person}{Matthew
  Hearn}, {and} \bibinfo{person}{Dawn Lawrie}.}
  \bibinfo{year}{2011}\natexlab{}.
\newblock \showarticletitle{Improving identifier informativeness using part of
  speech information}. In \bibinfo{booktitle}{{\em MSR}}.
\newblock


\bibitem[\protect\citeauthoryear{Campbell, Hindle, and Amaral}{Campbell
  et~al\mbox{.}}{2014}]%
        {campbell14}
\bibfield{author}{\bibinfo{person}{Joshua~Charles Campbell},
  \bibinfo{person}{Abram Hindle}, {and} \bibinfo{person}{Jos{\'e}~Nelson
  Amaral}.} \bibinfo{year}{2014}\natexlab{}.
\newblock \showarticletitle{Syntax errors just aren't natural: improving error
  reporting with language models}. In \bibinfo{booktitle}{{\em MSR}}.
\newblock


\bibitem[\protect\citeauthoryear{Cheon and Leavens}{Cheon and Leavens}{2002}]%
        {Cheon02aruntime}
\bibfield{author}{\bibinfo{person}{Yoonsik Cheon} {and}
  \bibinfo{person}{Gary~T. Leavens}.} \bibinfo{year}{2002}\natexlab{}.
\newblock \showarticletitle{A runtime assertion checker for the Java Modeling
  Language (JML)}. In \bibinfo{booktitle}{{\em Proceedings Of The International
  Conference On Software Engineering Research And Practice (SERP ’02), Las
  Vegas}}. \bibinfo{publisher}{CSREA Press}, \bibinfo{pages}{322--328}.
\newblock


\bibitem[\protect\citeauthoryear{Cleland-Huang, Gotel, and
  Zisman}{Cleland-Huang et~al\mbox{.}}{2012}]%
        {cleland2012software}
\bibfield{editor}{\bibinfo{person}{Jane Cleland-Huang}, \bibinfo{person}{Orlena
  Gotel}, {and} \bibinfo{person}{Andrea Zisman}} (Eds.).
  \bibinfo{year}{2012}\natexlab{}.
\newblock \bibinfo{booktitle}{{\em Software and systems traceability}}.
\newblock \bibinfo{publisher}{Springer}.
\newblock


\bibitem[\protect\citeauthoryear{Dagan, Roth, Sammons, and Zanzotto}{Dagan
  et~al\mbox{.}}{2013}]%
        {2013Dagan}
\bibfield{author}{\bibinfo{person}{Ido Dagan}, \bibinfo{person}{Dan Roth},
  \bibinfo{person}{Mark Sammons}, {and} \bibinfo{person}{Fabio~Massimo
  Zanzotto}.} \bibinfo{year}{2013}\natexlab{}.
\newblock \bibinfo{booktitle}{{\em Recognizing Textual Entailment: Models and
  Applications}}.
\newblock \bibinfo{publisher}{Morgan {\&} Claypool Publishers}.
\newblock


\bibitem[\protect\citeauthoryear{Dam, Tran, and Pham}{Dam
  et~al\mbox{.}}{2016}]%
        {dam16lstm}
\bibfield{author}{\bibinfo{person}{Hoa~Khanh Dam}, \bibinfo{person}{Truyen
  Tran}, {and} \bibinfo{person}{Trang Pham}.} \bibinfo{year}{2016}\natexlab{}.
\newblock \showarticletitle{A deep language model for software code}. In
  \bibinfo{booktitle}{{\em Workshop on Naturalness of Software, co-located with
  the 24th ACM SIGSOFT International Symposium on the Foundations of Software
  Engineering (FSE).}}
\newblock


\bibitem[\protect\citeauthoryear{Fluri, W{\"u}rsch, and Gall}{Fluri
  et~al\mbox{.}}{2007}]%
        {fluri2007code}
\bibfield{author}{\bibinfo{person}{Beat Fluri}, \bibinfo{person}{Michael
  W{\"u}rsch}, {and} \bibinfo{person}{Harald~C Gall}.}
  \bibinfo{year}{2007}\natexlab{}.
\newblock \showarticletitle{Do code and comments co-evolve? on the relation
  between source code and comment changes}. In \bibinfo{booktitle}{{\em WCRE}}.
\newblock


\bibitem[\protect\citeauthoryear{Fudaba, Oda, Akabe, Neubig, Hata, Sakti, Toda,
  and Nakamura}{Fudaba et~al\mbox{.}}{2015}]%
        {fudaba15}
\bibfield{author}{\bibinfo{person}{Hiroyuki Fudaba}, \bibinfo{person}{Yusuke
  Oda}, \bibinfo{person}{Koichi Akabe}, \bibinfo{person}{Graham Neubig},
  \bibinfo{person}{Hideaki Hata}, \bibinfo{person}{Sakriani Sakti},
  \bibinfo{person}{Tomoki Toda}, {and} \bibinfo{person}{Satoshi Nakamura}.}
  \bibinfo{year}{2015}\natexlab{}.
\newblock \showarticletitle{Pseudogen: A Tool to Automatically Generate
  Pseudo-code from Source Code}. In \bibinfo{booktitle}{{\em International
  Conference on Automated Software Engineering (ASE) Tool Demos}}.
\newblock


\bibitem[\protect\citeauthoryear{Goodfellow, Bengio, and Courville}{Goodfellow
  et~al\mbox{.}}{2016}]%
        {Goodfellow:book}
\bibfield{author}{\bibinfo{person}{Ian Goodfellow}, \bibinfo{person}{Yoshua
  Bengio}, {and} \bibinfo{person}{Aaron Courville}.}
  \bibinfo{year}{2016}\natexlab{}.
\newblock \bibinfo{booktitle}{{\em Deep Learning}}.
\newblock \bibinfo{publisher}{MIT Press}.
\newblock


\bibitem[\protect\citeauthoryear{Gu, Zhang, Zhang, and Kim}{Gu
  et~al\mbox{.}}{2016}]%
        {Gu2016deepapi}
\bibfield{author}{\bibinfo{person}{Xiaodong Gu}, \bibinfo{person}{Hongyu
  Zhang}, \bibinfo{person}{Dongmei Zhang}, {and} \bibinfo{person}{Sunghun
  Kim}.} \bibinfo{year}{2016}\natexlab{}.
\newblock \showarticletitle{Deep API Learning}. In \bibinfo{booktitle}{{\em
  SIGSOFT International Symposium on Foundations of Software Engineering}} {\em
  (\bibinfo{series}{FSE 2016})}. \bibinfo{publisher}{ACM},
  \bibinfo{address}{New York, NY, USA}, \bibinfo{pages}{631--642}.
\newblock


\bibitem[\protect\citeauthoryear{Gupta, Malik, Pollock, and
  Vijay-Shanker}{Gupta et~al\mbox{.}}{2013}]%
        {gupta2013part}
\bibfield{author}{\bibinfo{person}{Samir Gupta}, \bibinfo{person}{Sana Malik},
  \bibinfo{person}{Lori Pollock}, {and} \bibinfo{person}{K Vijay-Shanker}.}
  \bibinfo{year}{2013}\natexlab{}.
\newblock \showarticletitle{Part-of-speech tagging of program identifiers for
  improved text-based software engineering tools}. In \bibinfo{booktitle}{{\em
  ICPC}}.
\newblock


\bibitem[\protect\citeauthoryear{Hellendoorn and Devanbu}{Hellendoorn and
  Devanbu}{2017a}]%
        {hellendoorn17deep}
\bibfield{author}{\bibinfo{person}{Vincent Hellendoorn} {and}
  \bibinfo{person}{Premkumar Devanbu}.} \bibinfo{year}{2017}\natexlab{a}.
\newblock \showarticletitle{Are Deep Neural Networks the best choice for
  modeling Source Code?}. In \bibinfo{booktitle}{{\em Foundations of Software
  Engineering (FSE)}}.
\newblock


\bibitem[\protect\citeauthoryear{Hellendoorn and Devanbu}{Hellendoorn and
  Devanbu}{2017b}]%
        {hellendoorn2017fse}
\bibfield{author}{\bibinfo{person}{Vincent~J. Hellendoorn} {and}
  \bibinfo{person}{Premkumar Devanbu}.} \bibinfo{year}{2017}\natexlab{b}.
\newblock \showarticletitle{Are Deep Neural Networks the Best Choice for
  Modeling Source Code?}. In \bibinfo{booktitle}{{\em Foundations of Software
  Engineering (FSE)}}. \bibinfo{pages}{763--773}.
\newblock


\bibitem[\protect\citeauthoryear{Hindle, Barr, Su, Devanbu, and Gable}{Hindle
  et~al\mbox{.}}{2012}]%
        {hindle12naturalness}
\bibfield{author}{\bibinfo{person}{Abram Hindle}, \bibinfo{person}{Earl Barr},
  \bibinfo{person}{Zhendong Su}, \bibinfo{person}{Prem Devanbu}, {and}
  \bibinfo{person}{Mark Gable}.} \bibinfo{year}{2012}\natexlab{}.
\newblock \showarticletitle{On the "Naturalness" of software}.
\newblock In \bibinfo{booktitle}{{\em International Conference on Software
  Engineering (ICSE)}}.
\newblock


\bibitem[\protect\citeauthoryear{Hochreiter and Schmidhuber}{Hochreiter and
  Schmidhuber}{1997}]%
        {lstm}
\bibfield{author}{\bibinfo{person}{S. Hochreiter} {and} \bibinfo{person}{J.
  Schmidhuber}.} \bibinfo{year}{1997}\natexlab{}.
\newblock \showarticletitle{Long short-term memory}.
\newblock \bibinfo{journal}{{\em Neural Computation\/}} \bibinfo{volume}{9},
  \bibinfo{number}{8} (\bibinfo{year}{1997}), \bibinfo{pages}{1735--1780}.
\newblock


\bibitem[\protect\citeauthoryear{Ibrahim, Bettenburg, Adams, and
  Hassan}{Ibrahim et~al\mbox{.}}{2012}]%
        {ibrahim2012relationship}
\bibfield{author}{\bibinfo{person}{Walid~M Ibrahim}, \bibinfo{person}{Nicolas
  Bettenburg}, \bibinfo{person}{Bram Adams}, {and} \bibinfo{person}{Ahmed~E
  Hassan}.} \bibinfo{year}{2012}\natexlab{}.
\newblock \showarticletitle{On the relationship between comment update
  practices and software bugs}.
\newblock \bibinfo{journal}{{\em JSS\/}} (\bibinfo{year}{2012}).
\newblock


\bibitem[\protect\citeauthoryear{Jones, Harrold, and Stasko}{Jones
  et~al\mbox{.}}{2002}]%
        {tarantula}
\bibfield{author}{\bibinfo{person}{James~A Jones}, \bibinfo{person}{Mary~Jean
  Harrold}, {and} \bibinfo{person}{John Stasko}.}
  \bibinfo{year}{2002}\natexlab{}.
\newblock \showarticletitle{Visualization of test information to assist fault
  localization}. In \bibinfo{booktitle}{{\em International Conference on
  Software Engineering (ICSE)}}. ACM, \bibinfo{pages}{467--477}.
\newblock


\bibitem[\protect\citeauthoryear{Khamis, Witte, and Rilling}{Khamis
  et~al\mbox{.}}{2010}]%
        {khamis10automatic}
\bibfield{author}{\bibinfo{person}{Ninus Khamis}, \bibinfo{person}{Ren{\'e}
  Witte}, {and} \bibinfo{person}{Juergen Rilling}.}
  \bibinfo{year}{2010}\natexlab{}.
\newblock \showarticletitle{Automatic Quality Assessment of Source Code
  Comments: The JavadocMiner}. In \bibinfo{booktitle}{{\em Proceedings of the
  Natural Language Processing and Information Systems, and 15th International
  Conference on Applications of Natural Language to Information Systems}} {\em
  (\bibinfo{series}{NLDB'10})}. \bibinfo{pages}{68--79}.
\newblock


\bibitem[\protect\citeauthoryear{Knuth}{Knuth}{1984}]%
        {knuth1984literate}
\bibfield{author}{\bibinfo{person}{Donald~E Knuth}.}
  \bibinfo{year}{1984}\natexlab{}.
\newblock \showarticletitle{Literate programming}.
\newblock \bibinfo{journal}{{\it Comput. J.}} \bibinfo{volume}{27},
  \bibinfo{number}{2} (\bibinfo{year}{1984}), \bibinfo{pages}{97--111}.
\newblock


\bibitem[\protect\citeauthoryear{Knuth}{Knuth}{1992}]%
        {knuth1992literate}
\bibfield{author}{\bibinfo{person}{Donald~E Knuth}.}
  \bibinfo{year}{1992}\natexlab{}.
\newblock \showarticletitle{Literate programming}.
\newblock \bibinfo{journal}{{\em CSLI\/}}  \bibinfo{volume}{1}
  (\bibinfo{year}{1992}).
\newblock


\bibitem[\protect\citeauthoryear{Ko, Myers, Coblenz, and Aung}{Ko
  et~al\mbox{.}}{2006}]%
        {ko06exploratory}
\bibfield{author}{\bibinfo{person}{Andrew~J Ko}, \bibinfo{person}{Brad~A
  Myers}, \bibinfo{person}{Michael~J Coblenz}, {and} \bibinfo{person}{Htet~Htet
  Aung}.} \bibinfo{year}{2006}\natexlab{}.
\newblock \showarticletitle{An exploratory study of how developers seek,
  relate, and collect relevant information during software maintenance tasks}.
\newblock \bibinfo{journal}{{\em Software Engineering, IEEE Transactions on\/}}
  \bibinfo{volume}{32}, \bibinfo{number}{12} (\bibinfo{year}{2006}),
  \bibinfo{pages}{971--987}.
\newblock


\bibitem[\protect\citeauthoryear{LaToza, Venolia, and DeLine}{LaToza
  et~al\mbox{.}}{2006}]%
        {latoza06maintaining}
\bibfield{author}{\bibinfo{person}{Thomas~D LaToza}, \bibinfo{person}{Gina
  Venolia}, {and} \bibinfo{person}{Robert DeLine}.}
  \bibinfo{year}{2006}\natexlab{}.
\newblock \showarticletitle{Maintaining mental models: a study of developer
  work habits}. In \bibinfo{booktitle}{{\em International Conference on
  Software Engineering (ICSE)}}. ACM, \bibinfo{pages}{492--501}.
\newblock


\bibitem[\protect\citeauthoryear{Leavens and Cheon}{Leavens and Cheon}{2006}]%
        {Jml}
\bibfield{author}{\bibinfo{person}{G. Leavens} {and} \bibinfo{person}{Y.
  Cheon}.} \bibinfo{year}{2006}\natexlab{}.
\newblock \bibinfo{title}{Design by Contract with JML}.
\newblock   (\bibinfo{year}{2006}).
\newblock


\bibitem[\protect\citeauthoryear{Lin, Wang, Pang, Vu, Zettlemoyer, and
  Ernst}{Lin et~al\mbox{.}}{2017}]%
        {LinWPVZE2017:TR}
\bibfield{author}{\bibinfo{person}{Xi~Victoria Lin}, \bibinfo{person}{Chenglong
  Wang}, \bibinfo{person}{Deric Pang}, \bibinfo{person}{Kevin Vu},
  \bibinfo{person}{Luke Zettlemoyer}, {and} \bibinfo{person}{Michael~D.
  Ernst}.} \bibinfo{year}{2017}\natexlab{}.
\newblock \bibinfo{booktitle}{{\em Program synthesis from natural language
  using recurrent neural networks}}.
\newblock \bibinfo{type}{{T}echnical {R}eport} UW-CSE-17-03-01.
  \bibinfo{institution}{University of Washington Department of Computer Science
  and Engineering}, \bibinfo{address}{Seattle, WA, USA}.
\newblock


\bibitem[\protect\citeauthoryear{Maddison and Tarlow}{Maddison and
  Tarlow}{2014}]%
        {maddison2014}
\bibfield{author}{\bibinfo{person}{Chris~J Maddison} {and}
  \bibinfo{person}{Daniel Tarlow}.} \bibinfo{year}{2014}\natexlab{}.
\newblock \showarticletitle{Structured Generative Models of Natural Source
  Code}.
\newblock In \bibinfo{booktitle}{{\em International Conference on Machine
  Learning (ICML)}}. \bibinfo{pages}{649--657}.
\newblock


\bibitem[\protect\citeauthoryear{Manning, Surdeanu, Bauer, Finkel, Bethard, and
  McClosky}{Manning et~al\mbox{.}}{2014}]%
        {corenlp}
\bibfield{author}{\bibinfo{person}{Christopher~D. Manning},
  \bibinfo{person}{Mihai Surdeanu}, \bibinfo{person}{John Bauer},
  \bibinfo{person}{Jenny Finkel}, \bibinfo{person}{Steven~J. Bethard}, {and}
  \bibinfo{person}{David McClosky}.} \bibinfo{year}{2014}\natexlab{}.
\newblock \showarticletitle{The {Stanford} {CoreNLP} Natural Language
  Processing Toolkit}. In \bibinfo{booktitle}{{\em Association for
  Computational Linguistics (ACL) System Demonstrations}}.
  \bibinfo{pages}{55--60}.
\newblock


\bibitem[\protect\citeauthoryear{Marcus, Santorini, and Marcinkiewicz}{Marcus
  et~al\mbox{.}}{1993}]%
        {marcus93building}
\bibfield{author}{\bibinfo{person}{Mitchell~P. Marcus},
  \bibinfo{person}{Beatrice Santorini}, {and} \bibinfo{person}{Mary~Ann
  Marcinkiewicz}.} \bibinfo{year}{1993}\natexlab{}.
\newblock \showarticletitle{Building a Large Annotated Corpus of {E}nglish: The
  {P}enn {T}reebank}.
\newblock \bibinfo{journal}{{\em Computational Linguistics\/}}
  \bibinfo{volume}{19}, \bibinfo{number}{2} (\bibinfo{year}{1993}),
  \bibinfo{pages}{313--330}.
\newblock


\bibitem[\protect\citeauthoryear{McConnell}{McConnell}{2004}]%
        {mcconnell2004code}
\bibfield{author}{\bibinfo{person}{Steve McConnell}.}
  \bibinfo{year}{2004}\natexlab{}.
\newblock \bibinfo{booktitle}{{\em Code complete}}.
\newblock \bibinfo{publisher}{Pearson Education}.
\newblock


\bibitem[\protect\citeauthoryear{{Melis}, {Dyer}, and {Blunsom}}{{Melis}
  et~al\mbox{.}}{2017}]%
        {mellis17lstm}
\bibfield{author}{\bibinfo{person}{G. {Melis}}, \bibinfo{person}{C. {Dyer}},
  {and} \bibinfo{person}{P. {Blunsom}}.} \bibinfo{year}{2017}\natexlab{}.
\newblock \showarticletitle{{On the State of the Art of Evaluation in Neural
  Language Models}}.
\newblock \bibinfo{journal}{{\em ArXiv e-prints\/}} (\bibinfo{date}{July}
  \bibinfo{year}{2017}).
\newblock


\bibitem[\protect\citeauthoryear{Movshovitz-Attias and
  W.~Cohen}{Movshovitz-Attias and W.~Cohen}{2013}]%
        {movshovitz13natural}
\bibfield{author}{\bibinfo{person}{Dana Movshovitz-Attias} {and}
  \bibinfo{person}{William W.~Cohen}.} \bibinfo{year}{2013}\natexlab{}.
\newblock \showarticletitle{Natural Language Models for Predicting Programming
  Comments}. In \bibinfo{booktitle}{{\em ACL}}.
\newblock


\bibitem[\protect\citeauthoryear{Nguyen, Nguyen, and Nguyen}{Nguyen
  et~al\mbox{.}}{2013a}]%
        {nguyen13lexical}
\bibfield{author}{\bibinfo{person}{Anh~Tuan Nguyen},
  \bibinfo{person}{Tung~Thanh Nguyen}, {and} \bibinfo{person}{Tien~N. Nguyen}.}
  \bibinfo{year}{2013}\natexlab{a}.
\newblock \showarticletitle{Lexical Statistical Machine Translation for
  Language Migration}. In \bibinfo{booktitle}{{\em NIER}}.
\newblock


\bibitem[\protect\citeauthoryear{Nguyen, Nguyen, Nguyen, and Nguyen}{Nguyen
  et~al\mbox{.}}{2013b}]%
        {nguyen13statistical}
\bibfield{author}{\bibinfo{person}{Tung~Thanh Nguyen},
  \bibinfo{person}{Anh~Tuan Nguyen}, \bibinfo{person}{Hoan~Anh Nguyen}, {and}
  \bibinfo{person}{Tien~N Nguyen}.} \bibinfo{year}{2013}\natexlab{b}.
\newblock \showarticletitle{A statistical semantic language model for source
  code}. In \bibinfo{booktitle}{{\em FSE}}. \bibinfo{pages}{532--542}.
\newblock


\bibitem[\protect\citeauthoryear{Oda, Fudaba, Neubig, Hata, Sakti, Toda, and
  Nakamura}{Oda et~al\mbox{.}}{2015}]%
        {oda15}
\bibfield{author}{\bibinfo{person}{Yusuke Oda}, \bibinfo{person}{Hiroyuki
  Fudaba}, \bibinfo{person}{Graham Neubig}, \bibinfo{person}{Hideaki Hata},
  \bibinfo{person}{Sakriani Sakti}, \bibinfo{person}{Tomoki Toda}, {and}
  \bibinfo{person}{Satoshi Nakamura}.} \bibinfo{year}{2015}\natexlab{}.
\newblock \showarticletitle{Learning to Generate Pseudo-code from Source Code
  using Statistical Machine Translation}. In \bibinfo{booktitle}{{\em
  International Conference on Automated Software Engineering (ASE)}}.
\newblock


\bibitem[\protect\citeauthoryear{Padioleau, Tan, and Zhou}{Padioleau
  et~al\mbox{.}}{2009}]%
        {padioleau09listening}
\bibfield{author}{\bibinfo{person}{Yoann Padioleau}, \bibinfo{person}{Lin Tan},
  {and} \bibinfo{person}{Yuanyuan Zhou}.} \bibinfo{year}{2009}\natexlab{}.
\newblock \showarticletitle{Listening to Programmers - {T}axonomies and
  Characteristics of Comments in Operating System Code}. In
  \bibinfo{booktitle}{{\em International Conference on Software Engineering
  (ICSE)}}.
\newblock


\bibitem[\protect\citeauthoryear{Pascarella and Bacchelli}{Pascarella and
  Bacchelli}{2017}]%
        {Pascarella:2017}
\bibfield{author}{\bibinfo{person}{Luca Pascarella} {and}
  \bibinfo{person}{Alberto Bacchelli}.} \bibinfo{year}{2017}\natexlab{}.
\newblock \showarticletitle{Classifying Code Comments in Java Open-source
  Software Systems}. In \bibinfo{booktitle}{{\em Proceedings of the 14th
  International Conference on Mining Software Repositories}} {\em
  (\bibinfo{series}{MSR '17})}. \bibinfo{publisher}{IEEE Press},
  \bibinfo{address}{Piscataway, NJ, USA}, \bibinfo{pages}{227--237}.
\newblock
\showISBNx{978-1-5386-1544-7}


\bibitem[\protect\citeauthoryear{Raskin}{Raskin}{2005}]%
        {raskin}
\bibfield{author}{\bibinfo{person}{Jef Raskin}.}
  \bibinfo{year}{2005}\natexlab{}.
\newblock \showarticletitle{Comments Are More Important Than Code}.
\newblock \bibinfo{journal}{{\em ACM Queue\/}} \bibinfo{volume}{3},
  \bibinfo{number}{2} (\bibinfo{year}{2005}).
\newblock


\bibitem[\protect\citeauthoryear{Raychev, Bielik, and Vechev}{Raychev
  et~al\mbox{.}}{2016}]%
        {raychev16}
\bibfield{author}{\bibinfo{person}{Veselin Raychev}, \bibinfo{person}{Pavol
  Bielik}, {and} \bibinfo{person}{Martin Vechev}.}
  \bibinfo{year}{2016}\natexlab{}.
\newblock \showarticletitle{Probabilistic Model for Code with Decision Trees}.
  In \bibinfo{booktitle}{{\em OOPSLA}}.
\newblock


\bibitem[\protect\citeauthoryear{Spinellis}{Spinellis}{2010}]%
        {spinellis10}
\bibfield{author}{\bibinfo{person}{Diomidis Spinellis}.}
  \bibinfo{year}{2010}\natexlab{}.
\newblock \showarticletitle{Code Documentation}.
\newblock \bibinfo{journal}{{\em IEEE Software\/}} \bibinfo{volume}{27},
  \bibinfo{number}{4} (\bibinfo{year}{2010}), \bibinfo{pages}{18--19}.
\newblock


\bibitem[\protect\citeauthoryear{Steidl, Hummel, and Juergens}{Steidl
  et~al\mbox{.}}{2013}]%
        {steidl13quality}
\bibfield{author}{\bibinfo{person}{Daniela Steidl}, \bibinfo{person}{Benjamin
  Hummel}, {and} \bibinfo{person}{Elmar Juergens}.}
  \bibinfo{year}{2013}\natexlab{}.
\newblock \showarticletitle{Quality Analysis of Source Code Comments}. In
  \bibinfo{booktitle}{{\em Proceedings of the 21st IEEE Internation Conference
  on Program Comprehension (ICPC'13)}}.
\newblock


\bibitem[\protect\citeauthoryear{Sutskever, Vinyals, and Le}{Sutskever
  et~al\mbox{.}}{2014}]%
        {sutskever14}
\bibfield{author}{\bibinfo{person}{Ilya Sutskever}, \bibinfo{person}{Oriol
  Vinyals}, {and} \bibinfo{person}{Quoc Le}.} \bibinfo{year}{2014}\natexlab{}.
\newblock \showarticletitle{Sequence to Sequence Learning with Neural
  Networks}. In \bibinfo{booktitle}{{\em Advances in Neural Information
  Processing Systems (NIPS)}}.
\newblock


\bibitem[\protect\citeauthoryear{Tan, Yuan, Krishna, and Zhou}{Tan
  et~al\mbox{.}}{2007}]%
        {icomment}
\bibfield{author}{\bibinfo{person}{Lin Tan}, \bibinfo{person}{Ding Yuan},
  \bibinfo{person}{Gopal Krishna}, {and} \bibinfo{person}{Yuanyuan Zhou}.}
  \bibinfo{year}{2007}\natexlab{}.
\newblock \showarticletitle{/*Icomment: Bugs or Bad Comments?*/}. In
  \bibinfo{booktitle}{{\em Proceedings of Twenty-first ACM SIGOPS Symposium on
  Operating Systems Principles}} {\em (\bibinfo{series}{SOSP '07})}.
  \bibinfo{publisher}{ACM}, \bibinfo{address}{New York, NY, USA},
  \bibinfo{pages}{145--158}.
\newblock


\bibitem[\protect\citeauthoryear{Tan, Zhou, and Padioleau}{Tan
  et~al\mbox{.}}{2011}]%
        {tan2011acomment}
\bibfield{author}{\bibinfo{person}{Lin Tan}, \bibinfo{person}{Yuanyuan Zhou},
  {and} \bibinfo{person}{Yoann Padioleau}.} \bibinfo{year}{2011}\natexlab{}.
\newblock \showarticletitle{a{C}omment: mining annotations from comments and
  code to detect interrupt related concurrency bugs}. In
  \bibinfo{booktitle}{{\em ICSE}}.
\newblock


\bibitem[\protect\citeauthoryear{Tan, Marinov, Tan, and Leavens}{Tan
  et~al\mbox{.}}{2012}]%
        {tcomment}
\bibfield{author}{\bibinfo{person}{Shin~Hwei Tan}, \bibinfo{person}{D.
  Marinov}, \bibinfo{person}{Lin Tan}, {and} \bibinfo{person}{G.T. Leavens}.}
  \bibinfo{year}{2012}\natexlab{}.
\newblock \showarticletitle{@tComment: Testing Javadoc Comments to Detect
  Comment-Code Inconsistencies}. In \bibinfo{booktitle}{{\em Software Testing,
  Verification and Validation (ICST), 2012 IEEE Fifth International Conference
  on}}. \bibinfo{pages}{260--269}.
\newblock


\bibitem[\protect\citeauthoryear{Tu, Su, and Devanbu}{Tu et~al\mbox{.}}{2014}]%
        {localness-software}
\bibfield{author}{\bibinfo{person}{Zhaopeng Tu}, \bibinfo{person}{Zhendong Su},
  {and} \bibinfo{person}{Premkumar Devanbu}.} \bibinfo{year}{2014}\natexlab{}.
\newblock \showarticletitle{On the Localness of Software}. In
  \bibinfo{booktitle}{{\em Symposium on Foundations of Software Engineering
  (FSE)}}. \bibinfo{pages}{269--280}.
\newblock


\bibitem[\protect\citeauthoryear{White, Vendome, Linares-V\'asquez, and
  Poshyvanyk}{White et~al\mbox{.}}{2015}]%
        {white15deep}
\bibfield{author}{\bibinfo{person}{M. White}, \bibinfo{person}{C. Vendome},
  \bibinfo{person}{M. Linares-V\'asquez}, {and} \bibinfo{person}{D.
  Poshyvanyk}.} \bibinfo{year}{2015}\natexlab{}.
\newblock \showarticletitle{Toward deep learning software repositories}. In
  \bibinfo{booktitle}{{\em Working Conference on Mining Software
  Repositories}}.
\newblock


\bibitem[\protect\citeauthoryear{Wong, Liu, and Tan}{Wong
  et~al\mbox{.}}{2015}]%
        {wong2015clocom}
\bibfield{author}{\bibinfo{person}{Edmund Wong}, \bibinfo{person}{Taiyue Liu},
  {and} \bibinfo{person}{Lin Tan}.} \bibinfo{year}{2015}\natexlab{}.
\newblock \showarticletitle{Clo{C}om: Mining existing source code for automatic
  comment generation}. In \bibinfo{booktitle}{{\em SANER}}.
\newblock


\end{thebibliography}

\end{document}